# Unveiling intrinsic bulk photovoltaic effect in atomically thin ReS$_2$

*Maria Ramos[1]\*, Tanweer Ahmed[1], Bao Q. Tu[1], Eleni Chatzikyriakou[2], Lucía Olano-Vegas[1], Beatriz Martín-García[1,4], M. Reyes Calvo[3,4], Stepan S. Tsirkin[2,4], Ivo Souza[2,4], Félix Casanova[1,4], Fernando de Juan[4,5], Marco Gobbi[2,4]\*, Luis E. Hueso[1,4]\**

[1]CIC nanoGUNE BRTA, 20018 Donostia-San Sebastián, Basque Country, Spain
[2]Centro de Física de Materiales CSIC-UPV/EHU, 20018 Donostia-San Sebastián, Basque Country, Spain
[3]BCMaterials, Basque Center for Materials, Applications and Nanostructures, UPV/EHU Science Park, 48940 Leioa, Spain
[4]IKERBASQUE, Basque Foundation for Science, 48009 Bilbao, Basque Country, Spain
[5]Donostia International Physics Center, 20018 Donostia-San Sebastián, Basque Country, Spain

\*Correspondence to: m.ramos@nanogune.eu, marco.gobbi@ehu.eus,

l.hueso@nanogune.eu

ABSTRACT. The bulk photovoltaic effect (BPVE) offers a promising avenue to surpass the efficiency limitations of current solar cell technology. However, disentangling intrinsic and extrinsic contributions to photocurrent remains a significant challenge. Here, we fabricate high-quality, lateral devices based on atomically thin ReS$_2$ with minimal contact resistance, providing an optimal platform for distinguishing intrinsic bulk photovoltaic signals from other extrinsic photocurrent contributions originating from interfacial effects. Our devices exhibit large bulk photovoltaic performance with intrinsic responsivities of ~1 mA/W in the visible range, without the need for external tuning knobs such as strain engineering. Our experimental findings are supported by theoretical calculations. Furthermore, our approach can be extrapolated to investigate the intrinsic BPVE in other non-centrosymmetric van der Waals materials, paving the way for a new generation of efficient light-harvesting devices.

MAIN TEXT. The exploitation of high-order, nonlinear photocurrents could potentially overcome the fundamental efficiency limit of current solar cells based on p-n junctions (1) (2) (3). This limit, known as the Shockley-Queisser limit, is directly related to the active material's bandgap. It dictates a maximum theoretical efficiency of about 33.7% for a single junction under standard conditions (4) (5). While multi-junction solar cells can achieve higher efficiencies, they remain restricted by the bandgaps of their constituent materials, limiting the utilization of lower-energy parts of the spectrum. Nonlinear photocurrents, unconstrained by bandgap limitations, present a feasible alternative to unlock the full potential of the solar spectrum and surpass the efficiency limit, extending light utilization into the infrared and beyond (6) (7) (8) (9).

The proportionality relation between second-order photocurrents and an applied electric field (or light polarization) in a material is dictated by the nonlinear conductivity tensor $\sigma^{(2)}$, which is determined by the material's symmetry. Breaking inversion symmetry is essential for the emergence of second-order photocurrents, leading to what is commonly referred to as "bulk" photovoltaic effect (BPVE). These terms reflect the characteristics of this phenomenon: the generation of a net DC photocurrent upon illumination of a single, homogenous material at zero applied bias.

While the discovery of the BPVE originated from research on ferroelectric perovskite oxides (10) (11) (12) (13) (14) (15) (16) (17), these materials have not been considered a viable solution against traditional solar cell technology due to their low light-to-electrical power conversion efficiency. Alternatively, van der Waals (vdW) materials present straightforward strategies for breaking inversion symmetry, offering a new and exciting avenue for BPVE research (18) (19) (20) (21) (22) (23) (24) (25) (26) (27) (28). Especially, vdW polar materials have demonstrated large responsivities on the order of 1-10 mA/W when implemented in vertical device architectures (29) (30) (31) (32) (33).

Among various promising vdW materials, $ReS_2$ emerges as a particularly attractive candidate for BPVE exploration. While bulk $ReS_2$ is centrosymmetric (34) (35), numerous studies highlighted the polytypism in the material (36) (37). Different stacking orders of the distorted 1T crystal structure (known as 1T') can lead to broken symmetries, particularly in the few-layer limit of $ReS_2$ where stacking energetics can differ from that of bulk samples. This broken symmetry is crucial for the emergence of several nonlinear phenomena such as second harmonic generation (SHG), as demonstrated by Song *et al.* (38), and ferroelectricity, as confirmed by Wan *et al.* (39) in few-layer $ReS_2$. These findings suggest the potential for a significant BPVE in few-layer $ReS_2$. Earlier reports on BPVE in $ReS_2$ have observed photocurrents at zero applied bias only at grain boundaries in polycrystalline flakes (40), or solely at the crystalline edge of heterostructures formed by two $ReS_2$ flakes (41). Furthermore, Wang *et al.* (42) reported BPVE in single-crystalline $ReS_2$ flakes using a vertical device geometry.

Nonetheless, vertical geometries face challenges in the identification of the intrinsic BPVE. A key challenge in BPVE research lies in distinguishing nonlinear photocurrents from those generated at Schottky barriers within a device. This differentiation is crucial, as the Schottky barrier photovoltaic effect can dominate the overall photocurrent, questioning whether the responsivity has an intrinsic origin. Lateral geometries with a focused incoming beam enable spatial separation of intrinsic and extrinsic effects, facilitating detection of photocurrents from the pristine, active material without ambiguity. However, sizeable BPVE signals under normal incidence excitation in lateral devices using non-centrosymmetric transition metal dichalcogenides usually require the application of external tuning knobs, such as strain (43) (44).

In this work, we overcome the limitations found across the literature by employing an optimal device engineering technique, which enables to clearly identify the BPVE in thin, single-crystalline $ReS_2$ flakes without the need of external tuning knobs or vertical device architectures. Our approach utilizes a lateral device geometry where the $ReS_2$ layer is in contact with graphite electrodes and encapsulated between hexagonal boron nitride (hBN). This yields devices with minimal contact resistance, reducing electrode-interface photocurrents and providing an optimal platform for sensitive detection of intrinsic non-linear photocurrents. We demonstrate the BPVE in $ReS_2$ by studying the dependence of the observed photocurrent on the incoming light polarization and its spatial distribution across the device. Remarkably, our approach enables the detection of a large BPVE response in $ReS_2$, with intrinsic responsivities of ~1mA/W. These experimental values are seconded by our calculated photocurrent response based on a non-centrosymmetric bilayer structure. Additionally, our fabrication approach can potentially be extrapolated to the study of the BPVE in other non-centrosymmetric vdW materials with no perpendicular two-fold rotational axis, where in-plane photocurrents are generically allowed even at normal incidence.

A critical aspect of successfully measuring a BPVE signal is employing a device design that minimizes unwanted effects and maximizes the contribution of the "intrinsic", active material. To achieve this, we employ a device fabrication with the following considerations: First, a lateral device geometry is chosen to separate the BPVE from other potential photovoltaic effects originating at interface between different materials, which are typically present in devices with vertical geometries. Second, the $ReS_2$ active layer is sandwiched in between two thin hBN flakes to minimize surface defects and prevent any influence from the substrate. Third, few-layer graphite flakes are used as electrodes in contact with $ReS_2$ to minimize contact resistance. Fourth, a pick-up technique is employed during the device assembly to ensure clean interfaces between all materials. Further details about device fabrication are provided in section S1.

Fig. 1a-b summarizes the different vdW layers present in the devices through simplified sketches from both 3D and cross-sectional viewpoints, respectively. A bilayer-thick $ReS_2$ flake serves as the active channel, contacted by two thin graphite electrodes and encapsulated between top and bottom hBN flakes (see section S2 for $ReS_2$ thickness estimation). The vdW stack illustrated in Fig. 1a-b is placed onto a $SiO_2/Si^{++}$ substrate.

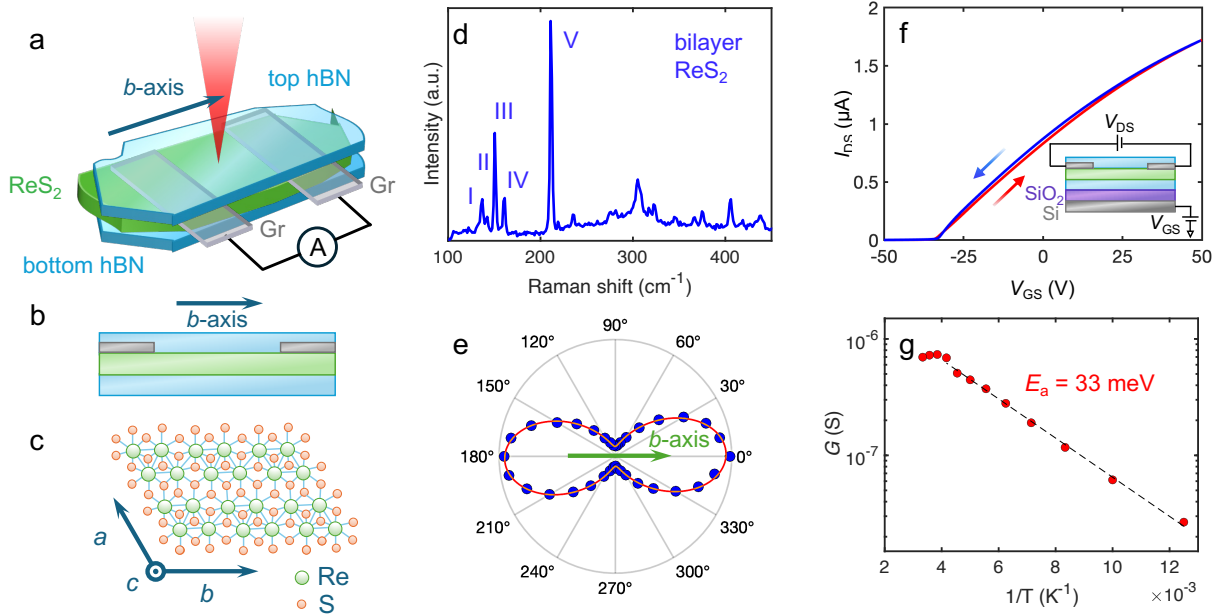

**Figure 1.** Simplified three-dimensional (a) and cross-sectional (b) device schemes showing the different vdW layers composing the system. The color coding chosen for each material layer in panel b is the same as for panel a. In panel a, a CW (continuous wave) laser is used to illuminate the device, while a current meter reads the photogenerated current. (c) Top-view atomic configuration of a monolayer ReS$_2$ with 1T' structure. (d) Raman spectrum of a bilayer ReS$_2$, where the first five vibrational modes can be clearly identified. (e) Polar plot of the A$_g$-like active Raman mode at ~212 cm$^{-1}$ (mode V in panel d). The orientation of the Re-Re bonds (*b*-axis) is identified along the maximum intensity. (f) Back- and forth-transfer curves of one of the fabricated devices, upon back-gate voltage (see sketch in the inset) and in dark conditions for $V_{DS}$ = 0.2 V. (g) Arrhenius plot for the conductance, where a linear fit to the data yields an activation energy of 33 meV.

The top-view atomic configuration of a monolayer ReS$_2$ with 1T' structure is shown in Fig. 1c, where Re-Re bonds form chains along the *b*-axis. The unit cell of ReS$_2$ contains four rhenium atoms and eight sulfur atoms, with the *a* and *b* axes forming an angle of 118° (34) (35). As indicated in panels a and b of Fig. 1, current detection is carried out along the *b*-axis of the ReS$_2$ flakes, as the crystals are naturally cleaved along this crystallographic direction. This is also confirmed through polarization-resolved Raman spectroscopy of the vibrational mode V (~212 cm$^{-1}$). As mode V involves out-of-plane vibrations of S atoms coupled with in-plane vibrations of Re atoms along the *b*-axis, a beam polarized along the direction of the Re atomic chain optimally couples with the in-plane vibrational component of the V mode.

Consequently, the polarization at which the intensity of mode V is maximum indicates the orientation of the *b*-axis (see Fig. 1d-e).

To ensure the quality and functionality of the fabricated devices, we check the transfer characteristics of a representative device (see Fig. 1f). These measurements reveal a well-defined on/off switching ratio of ~$10^3$ and a field-effect mobility of 12 cm$^2$/(V·s), indicating good quality and efficient carrier transport in agreement with previous reports (45). In addition, the transfer curves exhibit very small hysteresis, suggesting minimal charge trapping effects within the device.

Low contact resistance is essential for detecting second-order photocurrents. Our graphite electrodes minimize Schottky barriers, enabling efficient carrier extraction. This is confirmed by the low activation energy of our device, $E_a$ = 33 meV (see Fig. 1g), significantly lower than ReS$_2$ devices contacted to traditional metals (46) (47).

With the optimized ReS$_2$ devices in hand, we next investigate their intrinsic photovoltaic response. To do this, different $I_{DS}$-$V_{DS}$ curves in dark conditions and under illumination are studied. During the measurement, a linearly polarized laser emitting at 633 nm is used to illuminate the ReS$_2$ channel at different excitation powers. To avoid any photocurrent contribution from the ReS$_2$/graphite interface, the laser beam is tightly focused to a ~ 1 *µ*m spot size onto the center of the ReS$_2$ channel, whose length exceeds 5 *µ*m.

The current-voltage characteristics shown in Fig. 2a under dark conditions follow a linear trend, evidencing low contact resistance (see section S3 for larger applied voltages). Upon illumination, the $I_{DS}$-$V_{DS}$ curves exhibit a significant displacement that increases with the optical excitation power. This displacement indicates the generation of photocurrent at zero applied bias (short-circuit current, $I_{SC}$) and a non-zero voltage at zero current (open-circuit voltage, $V_{OC}$). These parameters define the operating regime of the device under illumination, where light-to-electrical power conversion occurs.

The values of the short-circuit current and the open-circuit voltage from Fig. 2a are shown respectively in Fig. 2b-c as a function of the excitation power. Both $I_{SC}$ and $V_{OC}$ exhibit a linear dependence on the excitation power ($P_{opt}$). This is indicative of a second-order photocurrent generation process, since power scales proportionally to the square of the electric field ($P_{opt} \propto |E|^2$).

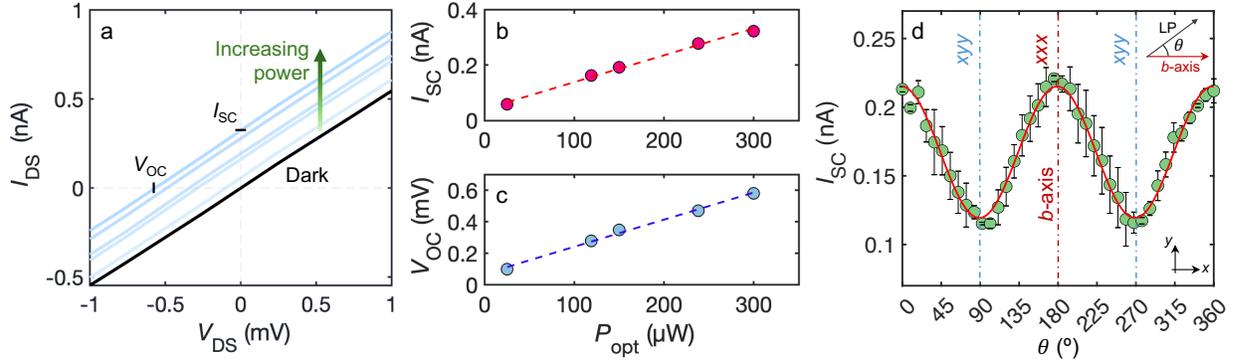

**Figure 2.** (a) $I_{DS}$-$V_{DS}$ characteristics of the ReS$_2$ device in the dark and upon illumination powers from 30 to 300 µW at 633 nm wavelength. (b) Short-circuit current extracted from panel a. (c) Open-circuit voltage obtained from panel a. (d) Angular dependence of the short-circuit current measured along the b-axis as a function of the linear polarization of the incoming light. The top right inset indicates the relative angle $\theta$ between the crystallographic b-axis of ReS$_2$ and the linear polarization of light (LP). Here, the b-axis is aligned with the x-coordinate (this happens at $\theta$ = 0º, 180º, 360º). Measurements are carried out with an excitation wavelength of 633 nm for an optical power of 120 mW.

A characteristic feature of a BPVE current is its sensitivity to the linear polarization of the incident light, whose response can be understood from the nonlinear conductivity $\sigma^{(2)}$ describing the material's response. To investigate this, we measured the short-circuit current at different angles of the linearly polarized light with respect to the b-axis of ReS$_2$. The results are shown in Fig. 2d, where the green dots represent the experimental data and the error bars indicate the standard deviation for each measurement. The results confirm a strong correlation between the short-circuit current and the light polarization, with a photocurrent approximately 2× larger when the light polarization aligns with the b-axis of ReS$_2$.

Interestingly, while the photocurrent exhibits a clear periodic modulation with light polarization, it remains non-zero for any polarization orientation. This agrees with previous observations of non-vanishing SHG for all polarization angles in ReS$_2$ (38) since these two effects are determined by the inherent crystal symmetry of ReS$_2$. Moreover, the non-zero mean photocurrent reveals a polar nature for ReS$_2$ crystals.

The results in Fig. 2d can be described by an equation capturing the relationship between a second-order photocurrent density along a given *i*-direction and the non-linear conductivity tensor as a quadratic electric field response:

$$j_i^{BPVE} = \sigma_{ijk}^{(2)} E_j E_k \quad (1)$$

where $\sigma_{ijk}^{(2)}$ is the second-order conductivity and $E_j$, $E_k$ are the components of the applied electric field along *j* and *k*, which in our case correspond to the electric field direction of the incoming light wave (or light polarization).

Since photocurrent is directly measured during experiments, from here onwards we provide the equations in terms of total photocurrent (*I*) and not photocurrent density (*j*). In our experiments, both the electric field (or light polarization) and the photocurrent detection are restricted to an in-plane configuration. Consequently, tensor components like those describing current flow along the *z*-axis can be disregarded. Also, since the photocurrent detection takes place along the *b*-axis of ReS$_2$, we adopt a convention where the *b*-axis aligns with the *x*-coordinate direction of our experimental setup. Therefore, the equation describing the photocurrent simplifies to:

$$I_x = t \cdot r \left( \sigma_{xxx}^{(2)} E_x^2 + \sigma_{xyy}^{(2)} E_y^2 + 2\sigma_{xxy}^{(2)} E_x E_y \right) \quad (2)$$

where *t* is the ReS$_2$ thickness and *r* is the beam radius. Considering an electric field amplitude $E_0$ forming an angle $\theta$ with respect to the *x*-axis, the electric field components can be written as $E_x = E_0 \cos\theta$ and $E_y = E_0 \sin\theta$. For a Gaussian beam, the electric field amplitude can be written in terms of optical power as $E_0^2 = 4P_{opt}/(nc\varepsilon_0 \pi r^2)$, where *n* is the refractive index of ReS$_2$, *c* is the speed of light and $\varepsilon_0$ is the vacuum permittivity. Substituting these expressions into eq. 2, the photocurrent as a function of the polarization angle becomes:

$$I_x = 4 \frac{t \cdot P_{opt}}{n\pi c \varepsilon_0 r} \left( \sigma_{xxx}^{(2)} \cos\theta^2 + \sigma_{xyy}^{(2)} \sin\theta^2 + 2\sigma_{xxy}^{(2)} \cos\theta \sin\theta \right) \quad (3)$$

A fit of the polarization dependent photocurrent in Fig. 2d using eq. 3 yields nonlinear conductivity values of $\sigma_{xxx}^{(2)}$ = (5.51 ± 0.07) mA/V², $\sigma_{xyy}^{(2)}$ = (3.06 ± 0.07) mA/V² and $\sigma_{xxy}^{(2)}$ = (-0.05

± 0.06) mA/V$^2$ within a 95% confidence interval. Further details on the fitting parameters used in this analysis are provided in section S4 of the Supporting Information.

To compare with other works, the responsivity of the device can be obtained from Fig. 2d. Since the responsivity is directly influenced by the sample dimension, a normalization based on sample size is needed for accurate comparisons across different studies. The "intrinsic" responsivity of our device is calculated from Fig. 2d as the ratio of the photocurrent density to the laser power density: $\kappa = (j^{BPVE}/I_{opt})$, where $j^{BPVE} = <I_{SC}>/(r \cdot t)$ and $I_{opt} = P_{opt}/(\pi \cdot r^2)$ being $<I_{SC}>$ the mean photocurrent, $r$ the laser beam radius, $t$ the ReS$_2$ thickness and $P_{opt}$ the optical power. For our bilayer device, this yields a BPVE intrinsic responsivity of 1.3 mA/W.

We employed scanning photocurrent microscopy (SPCM) to detect variations in the device's photovoltaic response across different regions. Fig. 3a shows an optical microscope image of a representative device with a four-layer ReS$_2$ active channel. The corresponding SPCM maps are shown in Fig. 3b-c, obtained under different illumination wavelengths and excitation powers, with zero applied bias. Both SPCM maps exhibit a remarkable similarity in the spatial distribution of the photogenerated current across the ReS$_2$ device for the two excitation wavelengths. The SPCM maps show a photocurrent with opposite sign at the lateral contact regions, located at both ends of the ReS$_2$ channel. The dominant mechanism of this photocurrent is a photovoltaic effect due to a built-in electric field at the graphite/ReS$_2$ interface. Here, the source (drain) barrier drives photogenerated electrons (holes) towards the semi-metal contact, resulting in a positive (negative) photocurrent. Furthermore, the SPCM maps also reveal a photocurrent generated within the ReS$_2$ channel.

The corresponding transverse profiles are respectively depicted in Fig. 3d-e, displaying a distinct and sharp increase in current when the ReS$_2$ active channel is illuminated. This rise indicates efficient photocurrent generation within the ReS$_2$ channel itself and its magnitude exhibits a direct proportionality to the illumination power used for excitation.

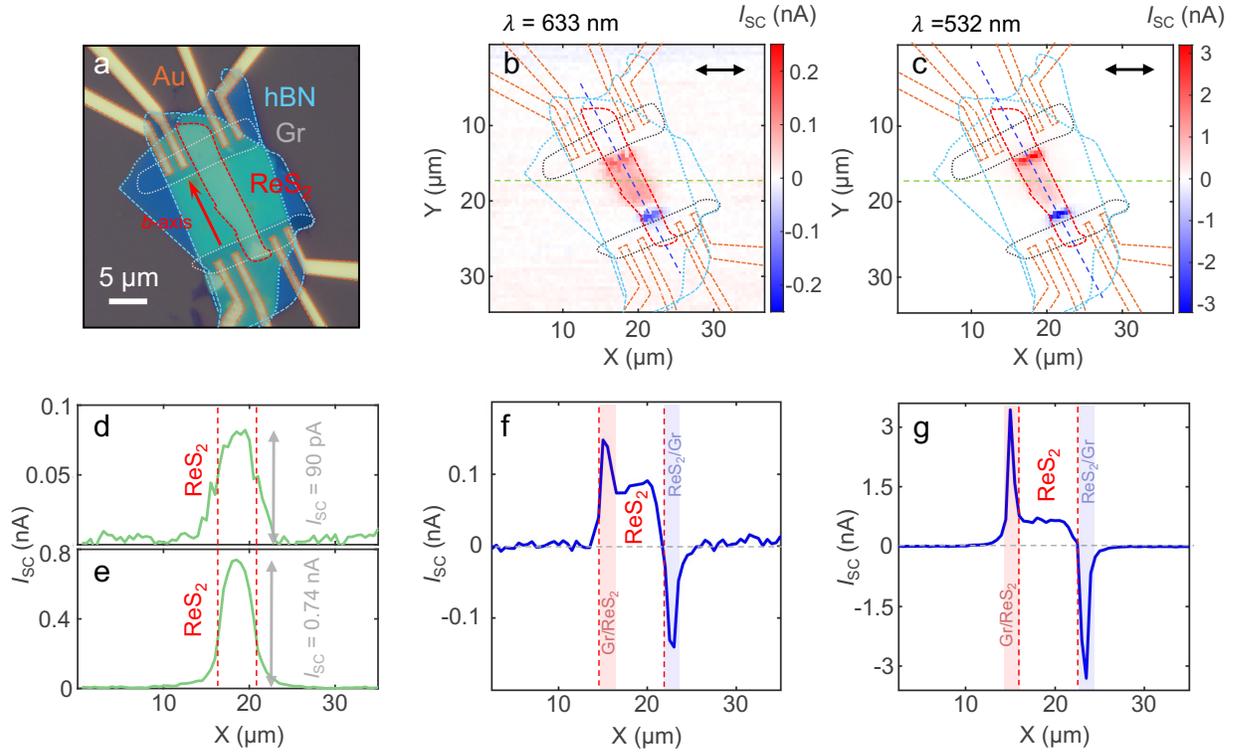

**Figure 3.** (a) Optical microscope image of a ReS$_2$ device. (b)-(c) Scanning photocurrent maps of the device shown in panel a for excitation wavelengths of 633 nm and 532 nm at excitation powers of 30 µW and 400 µW, respectively. The color scale represents the photocurrent intensity, with red and blue colors indicating positive and negative photocurrents, respectively. The black, double arrows at the top, right-hand side indicate the orientation of the linear polarization of light. (d)-(e) Transverse line profiles obtained from the photocurrent maps shown in panels b and c, respectively. (f)-(g) Longitudinal line profiles extracted from the photocurrent maps in panels b and c, respectively, where the reddish and blueish areas indicate positive and negative photocurrents built at the lateral Schottky barriers between the active channel and the graphite contacts.

While the transverse profiles focus on current fluctuations across the ReS$_2$ channel, the longitudinal profiles track the current along the ReS$_2$ active layer. The data obtained from the longitudinal line profiles in Fig. 3f-g provides valuable insights into the different contributions to the overall photocurrent signal. A constant dark current or noise level can be identified when the device is illuminated in non-photoresponsive regions. The photocurrent built at the lateral semi-metal/semiconductor interface appears at the edges of the ReS$_2$ channel with opposite sign contributions, with an extension of ~2-mm. Finally, in the central area of the device, the bulk photovoltaic current generated along the ReS$_2$ channel can be distinguished.

Significantly, the longitudinal line profiles reveal a disparity in the scaling behavior between the photovoltaic current built at the lateral semi-metal/semiconductor interface and that originating from the pristine $ReS_2$ flake, supporting the distinct origins of the two signals. Moreover, the magnitude of the photocurrent originated along the pristine $ReS_2$ flake in the longitudinal line profiles is consistent with the photocurrent intensity observed in the transverse line profiles. The uniformity of this signal across the SPCM maps and its sharp rise in the line profiles further support the dominance of the BPVE as the mechanism for photocurrent generation in our device.

Interestingly, we are not able to discern any BPVE signal in lateral $ReS_2$ devices fabricated with traditional metal contacts, such as prepatterned Ti/Au electrodes, as confirmed by the absence of any photocurrent at zero bias originated at the pristine $ReS_2$ channel through SPCM maps. Moreover, the Schottky barrier photocurrent is an order of magnitude greater than that observed in our graphite-contacted devices (see section S5 of the Supporting Information).

The device characterized in Fig. 3 exhibits intrinsic responsivities of 1.6 and 1.0 mA/W at wavelengths of 633 and 532 nm, respectively, for a $ReS_2$ channel containing four layers. Our low-contact-resistance devices outperform similar lateral devices characterized under analogous experimental conditions, where unstrained non-centrosymmetric 3R- and 2H-$MoS_2$ exhibited negligible BPVE currents (43) (44).

While bulk $ReS_2$ is centrosymmetric (34) (35), different stacking orders of the 1T' crystal structure can lead to broken symmetries in the few-layer limit. Recent works have pointed out the existence of a bilayer stacking different from the bulk one, where the top monolayer is rotated by 180° with respect to the bottom one. This generically leads to a non-centrosymmetric structure which is ferroelectric (39) and displays SHG (48), being also expected to generate a BPVE current.

We have performed an *ab initio* calculation of the non-linear conductivity in the non-centrosymmetric state defined in Ref. (39), using Wannier functions derived from first principles simulations (49) (see section S1 for details).

The response tensor $\sigma^{(2)}_{ijk}$ was calculated from:

$$\sigma^{(2)}_{ijk}(0,\omega,-\omega) = -\frac{i\pi e^3}{4\hbar^2}\int[d\mathbf{k}]\sum_{n,m}f_{nm}(I^{abc}_{mn}+I^{acb}_{mn})[\delta(\omega_{mn}-\omega)+\delta(\omega_{nm}-\omega)] \quad (4)$$

where $I^{abc}_{kmn}$ is the product of the interband dipole matrix and its generalized derivative with respect to the crystal momentum, $f_{nm} = f_n - f_m$ are Fermi occupation factors and $\hbar\omega_{nm} = E_m - E_n$ is the difference between the energies of bands *n* and *m*.

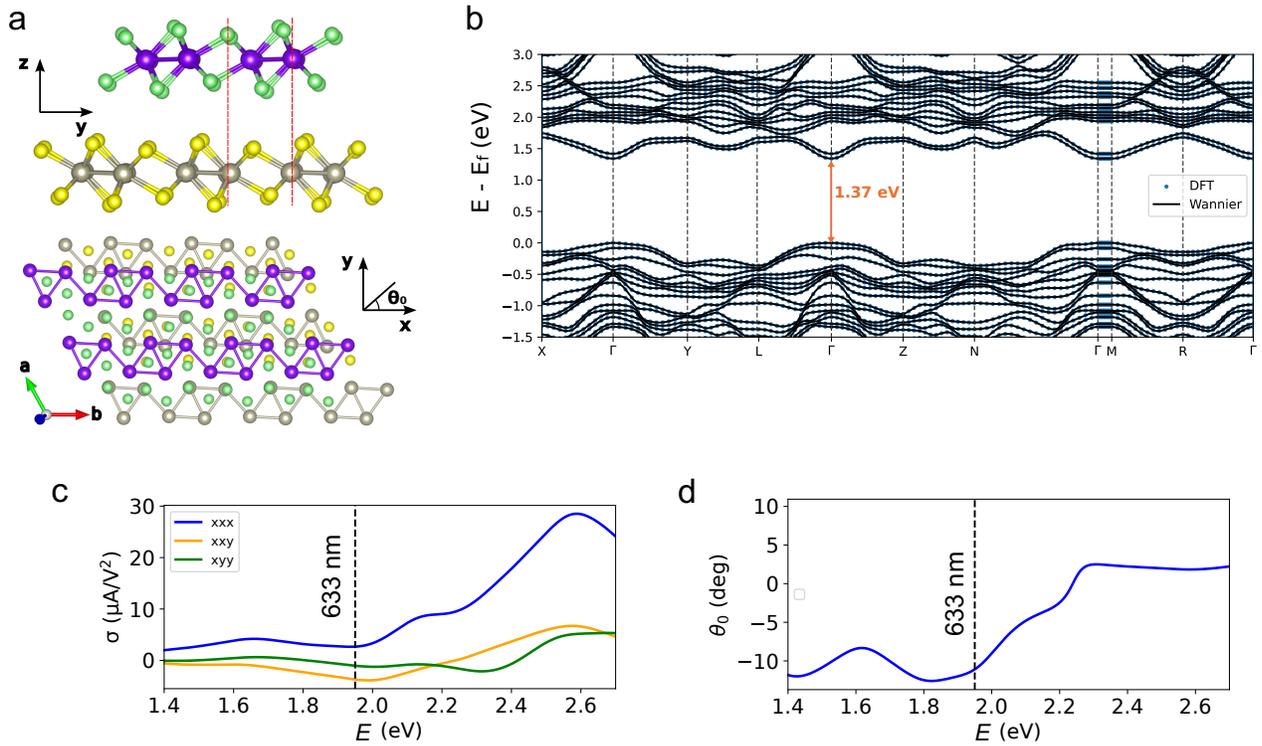

**Figure 4.** (a) Lattice structure of non-centrosymmetric ReS$_2$ bilayer. (b) Computed band structure comparing *ab-initio* and Wannierized bands. (c) Non-linear conductivity as a function of energy for all in-plane components. The dashed line corresponds to a laser wavelength of 633 nm (~ 1.95 eV) for comparison with experiment. (d) Angle of maximum photocurrent as a function of energy, which is within the range of [-12º, +2º] relative to the *b*-axis for all the calculated energies.

The results are depicted in Fig. 4, where the *x* coordinate corresponds to the *b*-axis of the ReS$_2$ crystal and to the direction of current flow in our experiments. At 633 nm wavelength, we obtained the components $\sigma^{(2)}_{xxx}$ = 2.68, $\sigma^{(2)}_{xyy}$ = -1.03 and $\sigma^{(2)}_{xxy}$ = -3.71 μA/V$^2$ (Fig. 4c). Our model predicts a photocurrent with the same order of magnitude as the one observed experimentally, something noteworthy in this field.

We obtained the angle $\theta_0$ at which photocurrent should be maximum from eq. 3 as: $\theta_0 = \frac{1}{2}\arctan 2\sigma^{(2)}_{xxy}/(\sigma^{(2)}_{xxx} - \sigma^{(2)}_{xyy})$. These results predict a maximum photocurrent at angles falling within the range of [-12º, +2º] with respect to the $b$-axis of ReS$_2$ for all the energies computed. These angles fall within an acceptable range of $\pm 10º$ angular error for our experimental setup, due to inherent limitations associated with aligning the polarization relative to the $b$-axis of the ReS$_2$ crystal.

In summary, our work unveils intrinsic BPVE in atomically thin ReS$_2$ through lateral device engineering. Scanning photocurrent microscopy and polarization-dependent photocurrent measurements provide compelling evidence for intrinsic photocurrent generation mechanisms. The $\sigma^{(2)}$ values extracted from experimental data are in the same order of magnitude as those obtained from theoretical calculations based on a non-centrosymmetric ReS$_2$ bilayer. Our device engineering strategy, combined with advanced characterization and theoretical calculations could be extended to study the intrinsic BPVE response in a broad range of non-centrosymmetric vdW materials.

ASSOCIATED CONTENT

**Supporting Information**. Methods; thickness estimation of ReS$_2$ active layers; $I_{SD} - V_{SD}$ characteristics up to 1 V; estimation of nonlinear conductivity from polarization-dependent photocurrent measurements; scanning photocurrent microscopy of a ReS$_2$ device with Ti/Au electrodes.

**Funding Sources**

ACKNOWLEDGMENT

This work has been funded by MCIN and by the European Union NextGenerationEU/PRTR-C17.I1, as well as by IKUR Strategy under the collaboration agreement between DIPC, CFM and nanoGUNE on behalf of the Department of Education of the Basque Government. This work received funding from MICIU/AEI/10.13039/501100011033 (Grant CEX2020-001038-M); from MICIU/AEI and ERDF/EU (Projects PID2021-128004NB-C21and PID2021-

122511OB-I00); and from MICIU/AEI and European Union NextGenerationEU/PRTR (Project PCI2021-122038-2A). B.M.G. and M.G. acknowledge support from MICIU/AEI and European Union NextGenerationEU/PRTR (Grants RYC2021-034836-I and RYC2021-031705-I). B.Q.T. acknowledges support from MICIU/AEI and ESF+ (Grant PRE2022-103674). L.O.V. thanks funding from Spanish MICIU/AEI/10.13039/501100011033 and ESF+ for the PhD grant PRE2022-104385. M.R.C. is funded through grant CNS2023-145151 funded by MCIN/AEI/10.13039/501100011033 and from European Union "NextGenerationEU"/PRTR). This work has been produced using the DIPC Supercomputing Center and the Aristotle University of Thessaloniki (AUTh) High Performance Computing Infrastructure and Resources. M.R. acknowledges Roger Llopis for the LabVIEW programs and Roger Llopis and Ralph Gay for the machining of setup components. T.A. acknowledges Roger Llopis and Ralph Gay for maintenance of CIC nanoGUNE's device fabrication facilities and financial supports from European Council via Grant agreement number 101046231 Fantasticof, and European Union via Marie Skłodowska-Curie grant agreement number 101107842 ACCESS.

ABBREVIATIONS

BPVE, bulk photovoltaic effect; CW, continuous wave; hBN, hexagonal boron nitride; vdW, van der Waals.

# Supporting Information

**Contents**

## 1. Methods

**Device fabrication.** Commercially available ReS$_2$ and hBN crystals were obtained from HQ Graphene, while the graphite crystals were sourced from NGS Naturgraphit GmbH. Each type of crystal was separately exfoliated on freshly cleaned SiO$_2$/Si substrates using the mechanical cleavage technique. ReS$_2$ flakes with a thickness of 2-10 layers and graphite flakes with a thickness of 4-5 layers were identified via optical contrast, and hBN flakes with a thickness of 20-40 nm were also identified. To fabricate the encapsulated graphene-contacted ReS$_2$ device, we used dry pick-up and transfer techniques as previously described in the literature (1-3). The exfoliation and dry transfer were performed inside an Ar-filled glovebox. We employed a drop of transparent LCC polymer (Lakme Colour Crush by Lakme India) as a sacrificial polymer (SP) layer. The convex meniscus of the SP was used to pick up the top hBN layer. Subsequently, two graphite layers were picked up one by one beneath the top hBN layer, aligned parallel to each other with approximately 10 µm distance. Next, the ReS$_2$ and bottom hBN were picked up to complete the heterostructure. All pick-ups were performed at ~ 60°C. The SP, along with the heterostructure, was then placed on a freshly cleaned 300 nm p$^{++}$-SiO$_2$/Si chip at ~ 100°C. The chip was immersed in a 1:3 acetone: IPA solution overnight to dissolve the SP, leaving a clean encapsulated heterostructure on the chip. Standard e-beam lithography was used to design the electrical leads. Reactive ion etching (RIE) was applied to selected areas of the hBN-encapsulated graphite to metallize on the graphite contacts. The RIE process, utilizing SF$_6$ gas at a flow rate of 20 sccm, a pressure of 20 mTorr, and an RF power of 35 W for approximately 15 seconds, etched away the top hBN. It is worth noting that the SF$_6$ RIE recipe etches hBN 90 times faster than graphene, effectively stopping at the graphite layer. Following this, 5/50 nm Ti/Au electrodes were evaporated to create metal contacts to complete the device.

**µ-Raman spectroscopy.** A Renishaw in Via Qontor microscope was used to characterize the Raman spectra of the samples at room temperature using a 532-nm CW laser as the excitation source. A 50× focusing objective providing a laser spot size of ~1 µm was used during data acquisition. To investigate the polarization dependence of the Raman modes, the samples were rotated relative to the polarization direction of the incident laser.

**Atomic force microscopy.** To study the thickness and topography of the devices, an AFM microscope (Bruker) was used in non-contact mode with a standard tip and in ambient conditions. Scanning speeds equal to or lower than 0.6 lines/s were used during acquisition for better quality imaging.

**Photocurrent measurements and scanning photocurrent microscopy.** Photocurrent measurements were performed as an extended home-built setup of a Renishaw in Via Qontor Raman microscope. The instrument's 532-nm and 633-nm CW excitation sources were employed for photoexcitation within a power range of 10 – 300 µW. The devices were mounted onto a commercial cryostat (Linkam, HFS350EV-PB4) provided of an optical window and electrical connections. The optical window provided both Raman spectroscopy and photocurrent measurements simultaneously. Electrical contacts were established via wire bonding for photocurrent detection using a Keithley 2634B sourcemeter. To study the polarization dependence of the photocurrent, a half-wave plate was inserted in the optical path prior to the focusing objective. The half-wave plate was then rotated relative to the incident laser polarization. Scanning photocurrent microscopy was conducted using the same experimental setup. During this measurement, a tightly focused laser beam (~ 1 µm diameter) was used to illuminate the different areas of the device while scanning the sample and the photocurrent generated at each position was measured. All photocurrent measurements were conducted in ambient conditions.

**Activation energy.** The activation energy of the devices was determined from temperature-dependent $I_{DS}$-$V_{DS}$ measurements in the temperature range of 80 to 300 K, with a step size of 20 K (see Fig. S1), and with a stabilization time of 12 minutes before each measurement. To do this, the devices were placed in a commercial cryostat (Linkam, HFS350EV-PB4) with a liquid nitrogen flux and using the same electronic components as for photocurrent measurements. The activation energy was then extracted from the variation of the channel conductance with temperature at $V_{DS}$ = 0.1 V, which is given by the equation $G(T) = G_0 e^{-E_a/K_b T}$, corresponding to the slope of the Arrhenius plot.

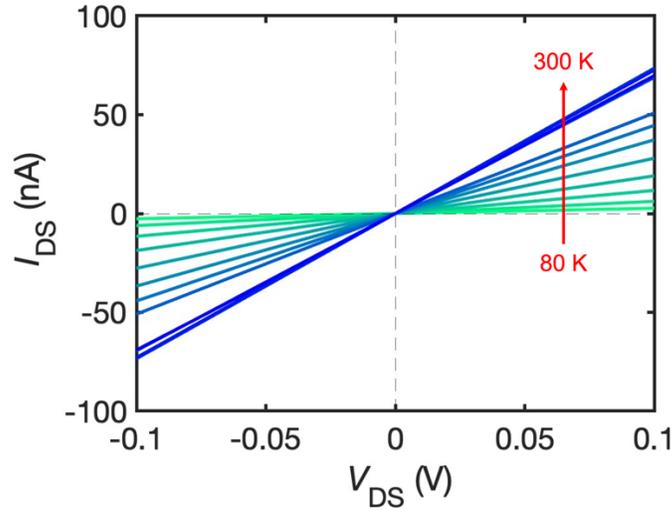

**Figure S5.** Temperature-dependent $I_{DS} - V_{DS}$ curves of a bilayer ReS$_2$ device, in the temperature range of 80 to 300 K with a step size of 20 K.

First principles simulations. Since in the 1T' structure there is an inversion center at the center of the 4-Re cluster, stacking two identical copies with an arbitrary lateral shift between them leads to a new inversion center exactly in between the original (intralayer) inversion centers.

The simulated structure was derived from the AB stacking phase of bilayer ReS$_2$ (4), where each monolayer has inversion symmetry, and the top layer is rotated 180º with respect to the center of inversion on the xy plane, while the atoms of the top layer are displaced on the same plane to the non-inversion symmetric minimum phase (5).

ReS$_2$ was simulated using QuantumEspresso (6-7). Non-spin-polarized simulations were performed. Optimized, norm-conserving, scalar relativistic pseudopotential with a Perdew–Burke–Ernzerhof exchange-correlation functional were employed (8-9). A wavefunction cut-off 90 Ry was used with a *k*-point grid of 9 × 9 × 1. vdW interactions were considered using the DFT-D3 model. The bilayer, inversion symmetric structure was originally relaxed to the point that the forces on ions were below 2×10$^{-6}$ Ha/bohr (0.1 meV/Å). The resulting in-plane lattice constants are *a* = 6.49 and *b* = 6.38 Å. The distance between two Re atoms along the z axis was ~ 6.2 Å, which agrees with calculations of the bulk primitive structure, with the use of vdW interactions that resulted in a *c* lattice constant of 6.21 Å. Overall, these values are in good agreement with previous theoretical (10) and experimental results (11). The AB structure was then derived from this relaxed bilayer structure and the atoms of the top layer were displaced on the xy plane, so as to arrive approximately at the *A'* state, which is described in Ref. (5). A final relaxation step was performed, which slightly moved the atoms of the top layer to their minimum at the *A'* position (~ 0.1 Å on each cartesian direction). It was also ensured that in the relaxed structure the *b* lattice vector was aligned with the x axis.

Projected Wannier functions were constructed for the bilayer ReS$_2$ in state *A'* using Wannier90 (12). A frozen energy window was added for all valence bands and up to ~3 eV from the highest occupied valence band. The Re(d) and S(p) orbitals were used as the basis set. WannierBerri was used for the shift current calculations (13). A grid of 150 × 150 × 28 was used. The parameter used for the regularization procedure to avoid near degeneracies in the sum-over-states was set to $\eta$ = 0.04 eV and a fixed Lorentzian broadening width of 0.1 eV was used.

## 2. Thickness estimation of ReS$_2$ active layers

**μ-Raman spectroscopy**

μ-Raman spectroscopy was employed to accurately characterize the thickness of the atomically thin ReS$_2$ flakes composing our two devices exhibiting intrinsic BPVE. Previous studies have established that Raman spectroscopy can effectively distinguish the thickness of ReS$_2$, particularly for very few-layer structures, by analyzing the difference between vibrational modes III (~ 150 cm$^{-1}$) and I ((~ 136 cm$^{-1}$) (see Ref. (14)).

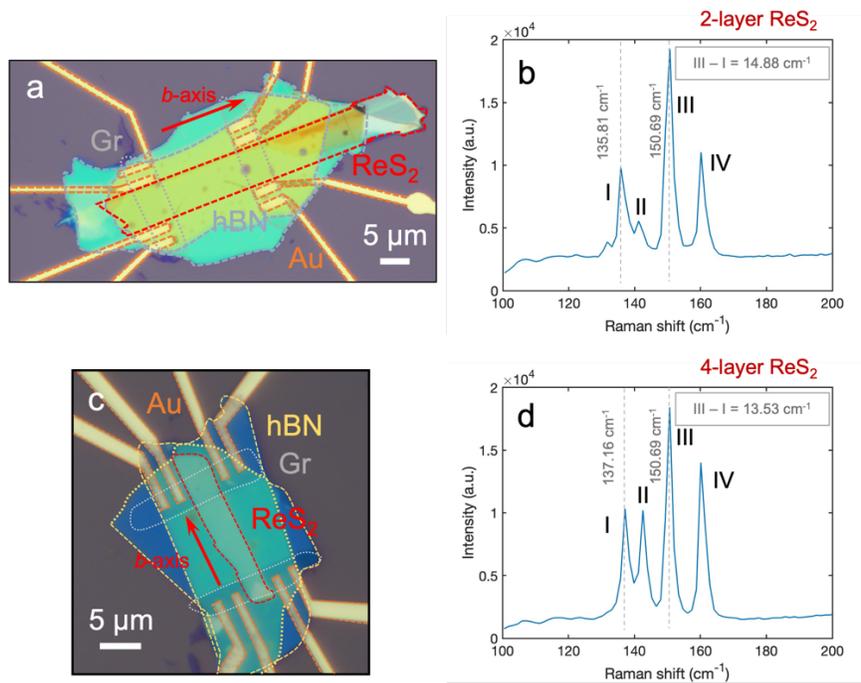

**Figure S6.** (a) Optical image of device *A*. (b) Raman spectrum of device *A* obtained from the active ReS$_2$ channel. The frequency difference between modes III and I is 14.88 cm$^{-1}$, in agreement with a bilayer ReS$_2$ flake. (c) Optical image of device *B*. (d) Raman spectrum of device *B* obtained from the center of the active ReS$_2$ channel. The difference between vibrational modes III and I is 13.53 cm$^{-1}$, revealing a four-layer ReS$_2$ flake.

Fig. S2 presents optical images of the two fabricated devices (device *A*, characterized in Fig. 2, and device *B*, characterized in Fig. 3 of the main manuscript) and representative Raman spectra for each of them within the frequency range of 100 – 200 cm$^{-1}$. For each device, three Raman spectra were acquired at different locations within the ReS$_2$ active channel. The calculated frequency difference between modes III and I remained constant for all the three spectra obtained from individual devices. The results are summarized in Fig. S2(b), (d).

By comparing these values with those reported in previous studies (14), we were able to accurately estimate the thickness of device *A* as a two-layer $ReS_2$ flake and device *B* as a four-layer $ReS_2$ flake.

**Optical color contrast**

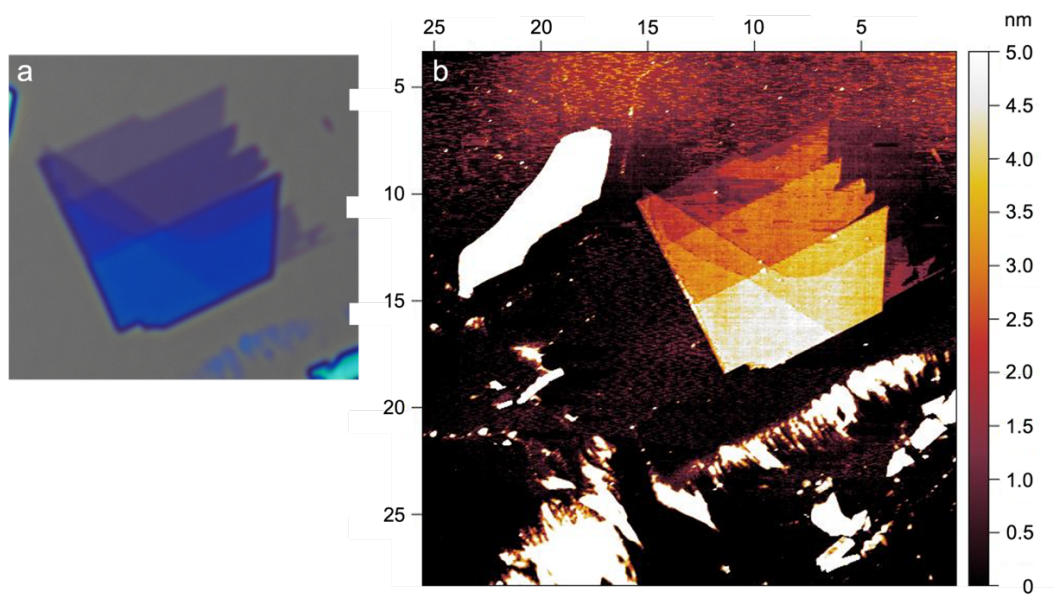

**Figure S7.** (a) Optical image of a $ReS_2$ flake with different thicknesses onto a $SiO_2$ substrate. (b) Atomic force microscope image of the flake shown in panel a.

**Table S1.** Optical color contrast and its corresponding number of ReS$_2$ layers obtained from the data in Fig. S3.

| Color | # of layers |
|---|---|
| 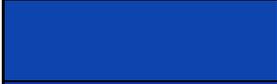 | 7 |
| 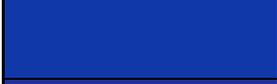 | 6 |
| 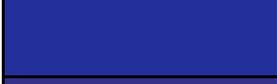 | 5 |
| 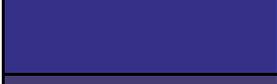 | 4 |
| 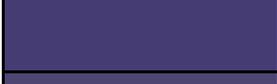 | 3 |
| 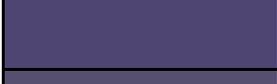 | 2 |
| 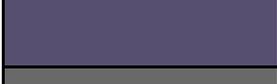 | 1 |
| 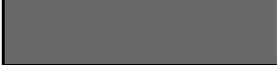 | SiO$_2$ |

Following the guideline in Table S1, we could have a rough estimation of the thickness of the desired ReS$_2$ flakes before proceeding with the device fabrication.

**Atomic force microscopy**

As a complementary thickness characterization technique, we performed atomic force microscopy (AFM) measurements of the two devices showing BPVE several months after their fabrication and BPVE characterization. During this time, the devices were exposed to air and the AFM measurements might be affected by the accumulation of dust and other contaminants on the surface. Additionally, the presence of the hBN encapsulation introduced challenges in accurately determining the thickness of the underlying ReS$_2$ flake, which can lead to an overestimation of the flake thickness in case of small air gaps or surface contaminants deposited after several months from device fabrication. Nevertheless, we provide these results in Fig. S4 and Fig. S5.

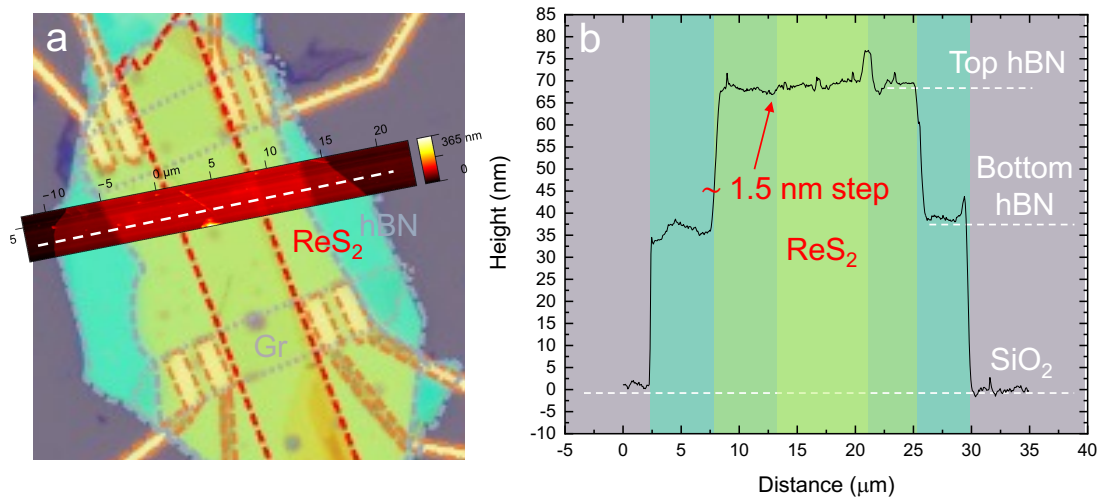

**Figure S8.** (a) Optical image of device *A* (characterized in Fig. 2 of the main manuscript) with an overlaid AFM image of the corresponding area. (b) Transverse line profile from the AFM image in panel a, extracted from the dashed white line. Here, the ReS$_2$ layer has a thickness of ~ 1.5 nm.

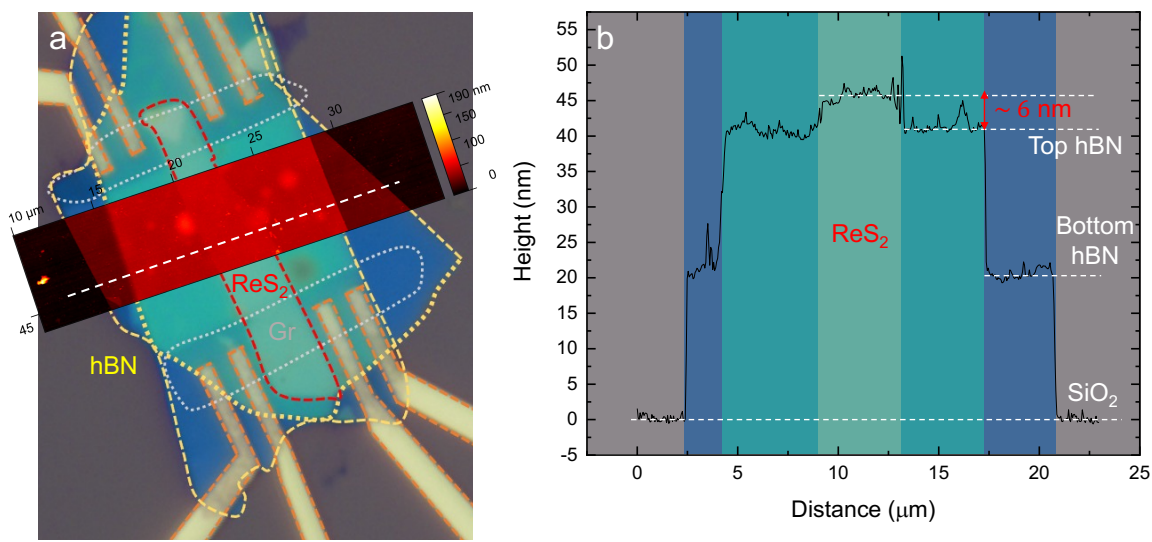

**Figure S9.** (a) Optical image of device *B* (characterized in Fig. 3 of the main text) with an overlaid AFM image of the corresponding device area. (b) Transverse device line profile extracted from the dashed white line in panel a. Here, the active ReS$_2$ channel has an approximated thickness of about 5 to 6 nm.

The thickness estimate obtained from AFM for the device in Fig. S5 does not align with the thickness derived from Raman spectroscopy (Fig. S2 (d)). Based on the consistency and accuracy of Raman spectroscopy in measuring ReS$_2$ thickness at its few-layer limit, we consider the Raman-derived values to be the most trustworthy.

## 3. $I_{SD}$ – $V_{SD}$ characteristics up to 1 V

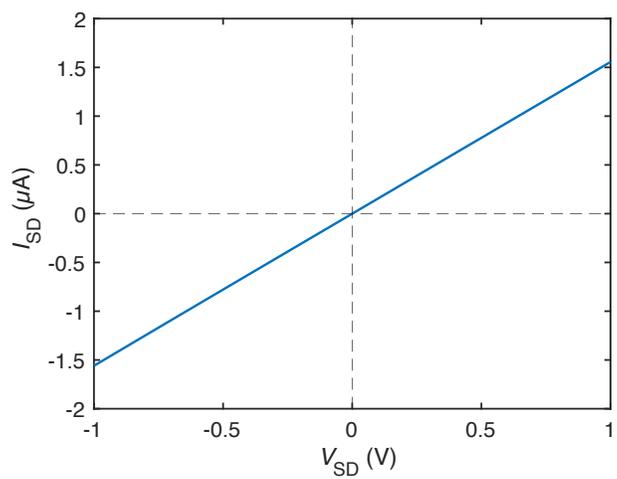

**Figure S6.** Output curve of a representative ReS$_2$ device with graphite contacts under dark conditions.

## 4. Estimation of nonlinear conductivity from polarization-dependent photocurrent measurements

Non-centrosymmetric materials can exhibit a second-order photocurrent response. Under linearly polarized light, a photocurrent density $j_i$ along the *i* direction can be generated:

$$j_i^{BPVE} = \sigma_{ijk}^{(2)} E_j E_k \qquad (S1)$$

Here, $\sigma_{ijk}^{(2)}$ is the nonlinear conductivity, and $E_j$, $E_k$ are the electric fields applied along *j* and *k* directions, which in our case correspond to the electric field of the incoming light wave or light polarization direction.

$\sigma_{ijk}^{(2)}$ is a third-rank tensor that captures the material's ability to conduct current in response to a second-order electric field dependence. Due to its tensorial nature, it can be expressed in a matrix form to account for all possible combinations of electric field components:

$$\begin{pmatrix} j_x \\ j_y \\ j_z \end{pmatrix} = \begin{pmatrix} \sigma_{xxx}^{(2)} & \sigma_{xyy}^{(2)} & \sigma_{xzz}^{(2)} & \sigma_{xyz}^{(2)} & \sigma_{xxz}^{(2)} & \sigma_{xxy}^{(2)} \\ \sigma_{yxx}^{(2)} & \sigma_{yyy}^{(2)} & \sigma_{yzz}^{(2)} & \sigma_{yyz}^{(2)} & \sigma_{yxz}^{(2)} & \sigma_{yxy}^{(2)} \\ \sigma_{zxx}^{(2)} & \sigma_{zyy}^{(2)} & \sigma_{zzz}^{(2)} & \sigma_{zyz}^{(2)} & \sigma_{zxz}^{(2)} & \sigma_{zxy}^{(2)} \end{pmatrix} \begin{pmatrix} E_x E_x \\ E_y E_y \\ E_z E_z \\ 2 E_y E_z \\ 2 E_x E_z \\ 2 E_x E_y \end{pmatrix} \qquad (S2)$$

Since few-layer ReS$_2$ breaks all symmetries, all the tensorial elements of $\sigma_{ijk}^{(2)}$ are non-zero. However, in our experiments, both the electric field (or light polarization) and the photocurrent detection are restricted to an in-plane configuration. Consequently, elements like those describing current flow along the *z*-axis can be disregarded. Also, since the photocurrent detection takes place along the *b*-axis of ReS$_2$, we adopt a convention where the *b*-axis aligns with the *x*-coordinate direction of our experimental setup. Therefore, the equation describing the photocurrent density simplifies to:

$$j_x = \sigma_{xxx}^{(2)} E_x^2 + \sigma_{xyy}^{(2)} E_y^2 + 2\sigma_{xxy}^{(2)} E_x E_y \qquad (S3)$$

Since photocurrent is directly measured during experiments, we can express eq. S3 in terms of photocurrent as:

$$I_x = t \cdot r \big(\sigma^{(2)}_{xxx} E_x^2 + \sigma^{(2)}_{xyy} E_y^2 + 2\sigma^{(2)}_{xxy} E_x E_y\big) \qquad (S4)$$

where $t$ is the thickness of the ReS$_2$ channel and $r$ is the laser beam radius. This approach assumes that the photocurrent is primarily generated within the area illuminated by the laser beam.

Considering an electric field amplitude $E_0$ forming an angle $\theta$ with respect to the x-axis, then $E_x$ and $E_y$ can be written as $E_x = E_0 \cos\theta$ and $E_y = E_0 \sin\theta$. Substituting these expressions into eq. S4, we obtain:

$$I_x = t \cdot r \cdot E_0^2 \big(\sigma^{(2)}_{xxx} \cos\theta^2 + \sigma^{(2)}_{xyy} \sin\theta^2 + 2\sigma^{(2)}_{xxy} \cos\theta \sin\theta\big) \qquad (S5)$$

For a monochromatic propagating wave, the light intensity is related to the electric field amplitude as:

$$I_{opt} = \tfrac{1}{2} c\varepsilon_0 n |E_0|^2 \qquad (S6)$$

where $c$ is the speed of light, $\varepsilon_0$ is the vacuum permittivity and $n$ is the refractive index of ReS$_2$.

Also, for a Gaussian beam with optical power $P_{opt}$ and beam radius $r$, the optical intensity can be written as:

$$I_{opt} = \frac{2 P_{opt}}{\pi r^2} \qquad (S7)$$

Substituting expressions S6 and S7 into eq. S5, then the photocurrent can be written as a function of the optical power:

$$I_x = 4 \frac{t \cdot P_{opt}}{n \pi c \varepsilon_0 r} \big(\sigma^{(2)}_{xxx} \cos\theta^2 + \sigma^{(2)}_{xyy} \sin\theta^2 + 2\sigma^{(2)}_{xxy} \cos\theta \sin\theta\big) \qquad (S8)$$

The previous equation has been used to fit the polarization-dependent photocurrent shown in Fig. 2d of the main text. The fitting was performed using the following parameter values: $t$ = 1.56 nm (corresponding to a bilayer ReS$_2$ active channel), $P_{opt}$ = 120 µW, $n$ = 4.6 (at 633 nm wavelength, obtained from Ref. (15)) and $r$ = 0.5 µm.

The nonlinear conductivity values for a bilayer ReS$_2$ obtained from this fit are: $\sigma^{(2)}_{xxx}$ = 5.51 μA/V², $\sigma^{(2)}_{xyy}$ = 3.06 μA/V² and $\sigma^{(2)}_{xxy}$ = -0.05 μA/V². The $R$-square value of the fit is 0.98.

## 5. Scanning photocurrent microscopy of a ReS₂ device with Ti/Au contacts

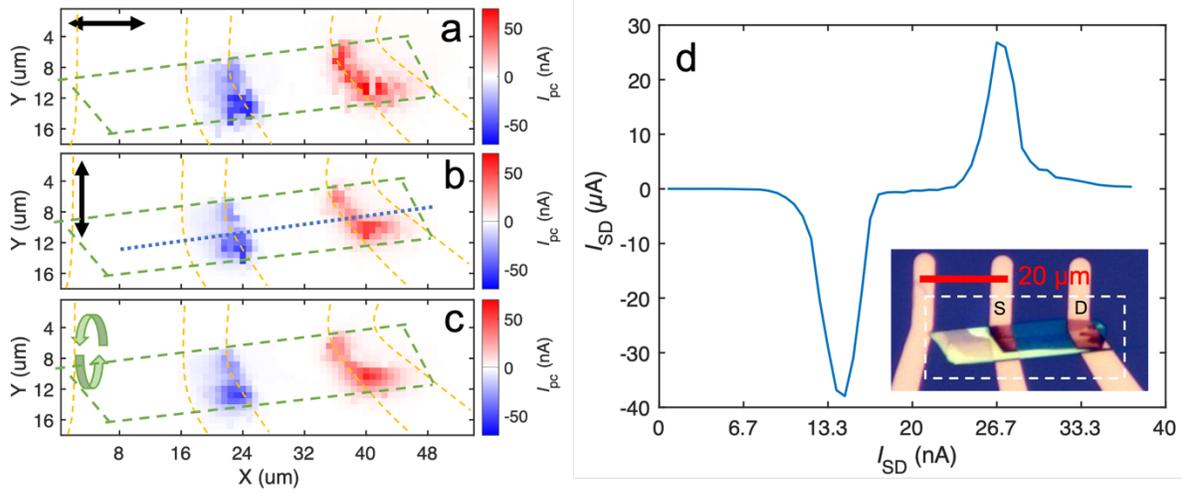

**Figure S7.** (a) – (c) Scanning photocurrent maps at 532-nm wavelength of a ReS$_2$ flake with different few-layer thicknesses and contacted to Ti/Au electrodes. The power used for all SPCM maps is 400 μW. The light polarization is indicated at the top-left corner of each map. (d) Longitudinal line profile obtained from the SPCM map shown in panel (b). The inset shows a microscopic image of the sample, where *S* and *D* denote the source and drain contacts.